\documentclass{iaus}
\usepackage{graphicx}

\title[The Real-Time Evolution of (V)LTP Objects]
{The Real-Time Evolution of Sakurai's Star (V4334 Sgr) and other (V)LTP Objects}

\author[van Hoof et al.]
{P.A.M. van Hoof$^1$, M.~Bryce$^{2}$, A.~Evans$^{3}$, S.P.S.~Eyres$^{4}$, \break M.~Hajduk$^{5,6}$,
 F.~Herwig$^{7}$, F.~Kerber$^{8}$, S.~Kimeswenger$^{9}$, \break J.A.~Lop\'ez$^{10}$, M.~Matsuura$^{5,11,12}$,
 D.L.~Pollacco$^{11}$, \break G.C. Van de Steene$^{1}$, \and A.A.~Zijlstra$^{5}$}

\affiliation{$^1$Royal Observatory of Belgium, Ringlaan 3, B-1180 Brussels, Belgium\\[\affilskip]
$^{2}$Jodrell Bank Observatory, University of Manchester, Macclesfield, Cheshire SK11 9DL, UK\\[\affilskip]
$^{3}$Department of Physics, Keele University, Staffordshire ST5 5BG, UK\\[\affilskip]
$^{4}$Centre for Astrophysics, University of Central Lancashire, Preston PRI 2HE, UK\\[\affilskip]
$^{5}$Univ.\ of Manchester, School of Physics \& Astronomy, P.O. Box 88, Manchester M60 1QD, UK\\[\affilskip]
$^{6}$Centrum Astronomii UMK, ul.\ Gagarina 11, PL-87-100 Toru\'n, Poland\\[\affilskip]
$^{7}$LANL, Theoretical Astrophysics Group T-6, MS B227, Los Alamos, NM 87545, USA\\[\affilskip]
$^{8}$European Southern Observatory, Karl-Schwarzschild-Strasse 2, 85478 Garching, Germany\\[\affilskip]
$^{9}$Inst.\ f{\"u}r Astro{--} \& Teilchenphysik, Univ.\ Innsbruck, Technikerstr. 25, 6020 Innsbruck, Austria\\[\affilskip]
$^{10}$Instituto de Astronom\'\i a UNAM, Apdo. Postal 877, 22800 Ensenada, BC, Mexico\\[\affilskip]
$^{11}$APS Division, Dept.\ of  Physics \& Astronomy, Queen's University Belfast, BT7 1NN, UK\\[\affilskip]
$^{12}$ARC, National Astronomical Observatory of Japan, Mitaka, Tokyo 181-8588, Japan}

\pubyear{2006}
\volume{234}
\jname{Planetary Nebulae in our Galaxy and Beyond}
\editors{M. J. Barlow \& R. H. M\'endez, eds.}
\pagerange{119--126}
\date{?? and in revised form ??}
\begin{document}

\maketitle

\begin{abstract}
We report on the progress of our on-going campaign to monitor the evolution of
the VLTP objects V4334 Sgr and V605 Aql, as well as the suspected (V)LTP
object CK Vul. V4334 Sgr does not show signs of increased ionization compared
to our previous observations in 2004. We obtained the first radio detection of
V605 Aql, indicating a strong increase in radio flux since 1987. We also
present the first radio detection of CK Vul and discuss the expansion of the
material ejected during the 1670 event.\keywords{stars: AGB and post-AGB,
stars: evolution, circumstellar matter, stars: winds, outflows}
\end{abstract}

\firstsection

\section{Introduction}

Helium shell flashes punctuate the evolution of a star on the thermal pulsing
Asymptotic Giant Branch (AGB). They occur whenever sufficient helium has
accumulated from the preceding phase of hydrogen burning. Following the helium
shell flash, a short phase of quiescent helium burning occurs, after which
hydrogen burning resumes. Nearly all of these thermal pulses occur near the
tip of the AGB, but approximately 20\% of all low and intermediate mass stars
are expected to suffer a final shell flash after they have left the AGB. If
the event occurs while the hydrogen burning shell is still active, ingestion
of the remaining hydrogen from the envelope into the helium burning shell is
impossible and the event is called a Late Thermal Pulse (LTP). However, if the
star is already on the cooling track and the hydrogen burning shell has
ceased, the remaining envelope can be ingested, leading to additional rapid
hydrogen-driven burning. This can result in an extremely fast real-time
evolution following a double-loop structure in the Hertzsprung-Russell diagram
(Herwig 2001, Hajduk \etal\ 2005). Such an event is called a Very Late Thermal
Pulse (VLTP). A recent review of this process can be found in Herwig (2005).
Very few VLTP events have been observed: only V4334 Sgr (Sakurai's Object) and
V605 Aql were discovered during their high-luminosity phase (respectively in
1996 and 1918). CK Vul (in 1670) is suspected to represent a third case (Evans
\etal\ 2002) and may either be an LTP or VLTP event. We are currently involved
in an effort to monitor the evolution of these objects. We presented results
of this campaign for Sakurai's Object in Hajduk \etal\ (2005). Below we will
give a brief update on the status of this campaign. We will also present first
results for V605 Aql and CK Vul.

\section{V4334 Sgr (Sakurai's Object)}

V4334 Sgr was the core of a previously unknown planetary nebula (PN) when it
erupted in 1992. In 2004 we obtained [O\,{\sc iii}] images using FORS1 on the
VLT, and 8.6 GHz radio data with the VLA. The old PN has been detected in both
data sets. This nebula is still visible because the O$^{2+}$ ions did not have
time yet to recombine. Hence the nebula has a ``memory'' of conditions before
the VLTP. Analysis of this spectrum showed that the old PN had already entered
upon the cooling track (Kerber \etal\ 1999, Pollacco 1999). During the VLTP
event, the central star ingested part of the surface material into deeper
layers where it was burned. The remainder has been ejected, exposing the
hydrogen-deficient layers below. The expelled material is now moving away from
the star at a velocity of $\sim$ 350~km\,s$^{-1}$ (Kerber \etal\ 2002) and
will eventually form a new hydrogen-deficient planetary nebula. Initial
spectra of the ejecta showed strong C$_2$ absorption bands, indicative of a
carbon-rich molecular chemistry (Asplund \etal\ 1997), but no dust. Later dust
formation did start, leading to significant circumstellar absorption. The
central star is now very faint (V=23 mag). The new ejecta have been detected
in [N\,{\sc ii}] and [O\,{\sc ii}] lines (Kerber \etal\ 2002), as well as
radio emission (Hajduk \etal\ 2005).

During 2005 we obtained new VLA data at 5 and 8~GHz in the AnB configuration.
This configuration achieves higher angular resolution than our earlier
observations. At 8~GHz we detected the source with a flux of 80 $\pm$
30~$\mu$Jy, in agreement with the 2004 observations. The deconvolved FWHM of
the object is 1.0 $\times$ 0.5~arcsec, which is smaller than reported in
Hajduk \etal\ (2005). The suspected bipolar morphology could not be confirmed.
We did not detect Sakurai's object at 5~GHz with a 3$\sigma$ upper limit of
50~$\mu$Jy/beam. This indicates that the radio emission is optically thick at
5~GHz. Assuming $T_{\rm e}$ = 4500~K (Hajduk \etal\ 2005) and a filling factor
of unity this yields a diameter $\Theta \leq 20$~mas, which is much smaller
than observed. This would imply a low filling factor, indicating the presence
of a thin shell or a very clumpy medium.

In 2005 we obtained new optical spectra with FORS1 on the VLT. Full analysis
of the spectra is not complete yet, but first indications are that there are
no big changes w.r.t. to the Kerber \etal\ (2002) spectrum, although changes
in the individual velocity components do appear to be present. In April 2005
we also obtained an infrared spectrum with IRS aboard the Spitzer Space
Telescope (Evans \etal\ 2006). The spectrum shows absorption bands of HCN and
polyyne molecules, but no atomic emission lines. In particular the [Ne\,{\sc
ii}] 12.8~$\mu$m line has not been detected. This is one of the first lines
that is expected to appear when reheating of the central star progresses.

\section{V605 Aql}

This star is sometimes called the ``older twin'' of Sakurai's Object. It
erupted around 1917. It is the central star of an older planetary nebula
(A58). The star is still very reddened and faint (V = 23~mag). It has a
Wolf-Rayet type spectrum (Seitter 1987, Clayton \etal, these proceedings).

We observed V605 Aql with the VLA on 4 April 2005. The flux was 440 $\pm$
50~$\mu$Jy at 5~GHz in the B configuration. The deconvolved FWHM was 0.7
$\times$ 0.3~arcsec, consistent with optical observations (e.g., Bond \&
Pollacco 2002). Previous observations by Rao \etal\ (1987) using the VLA only
yielded upper limits of 150~$\mu$Jy at 5~GHz, and 370~$\mu$Jy at 15~GHz. Hence
the radio flux has increased considerably since 1987. This could be due to a
decrease in optical depth (expansion of the ejecta) or an increase in emission
measure (e.g. due to increased stellar temperature). The latter is suggested
by changes in the spectrum of the core (Kimeswenger 2003, see also Clayton
\etal, these proceedings).

\section{CK Vul}

This object was first observed in 1670. Its nebula shows highly reddened
emission lines of H\,{\sc i}, [N\,{\sc ii}], [O\,{\sc iii}], and [S\,{\sc
ii}], but the central star has not been found yet. This, and other unusual
properties, make it unlikely that CK Vul is an ordinary nova. The observations
better fit the hypothesis that this is an LTP or VLTP object (Evans \etal\
2002).

\begin{figure}
\centerline{\includegraphics[height=8.5cm]{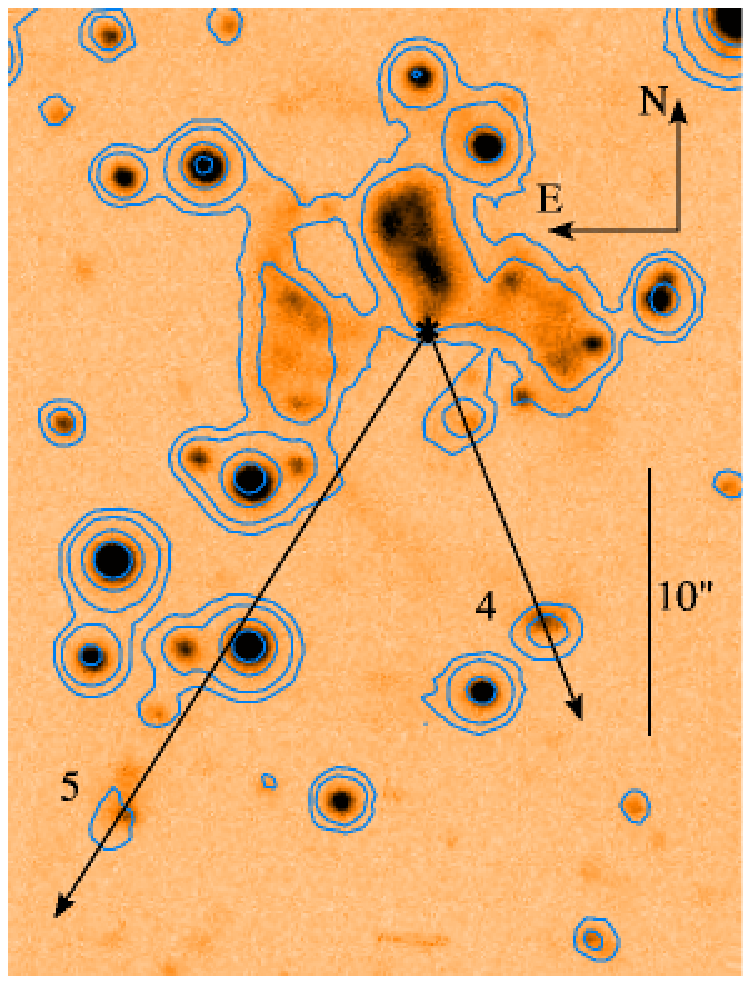}\hspace{1.1mm}\includegraphics[height=8.5cm]{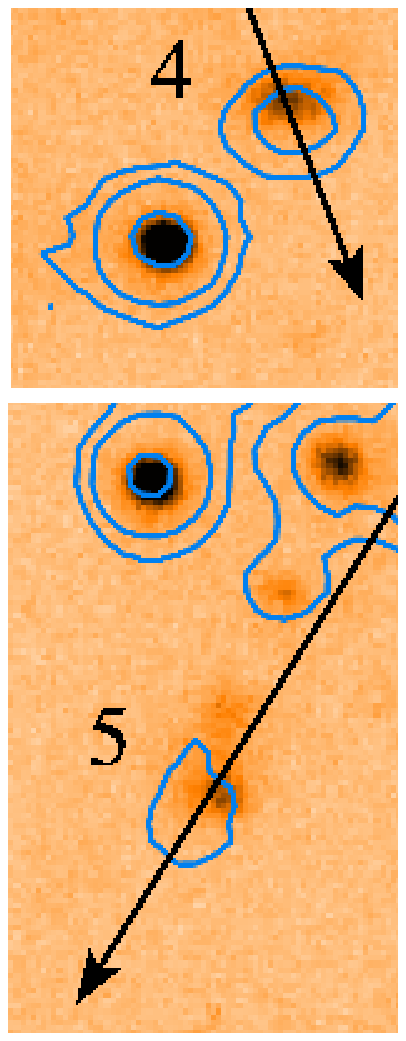}}
\caption{The expansion of the ejecta from the 1670 event in CK Vul. The
negative image shows H$\alpha$ observations from 1991 (Naylor \etal\ 1992),
the contours are H$\alpha$ observations from 2004 taken from the IPHAS survey
(Drew \etal\ 2005). The asterisk marks the position of the radio source. Knots
4 \& 5 are enlarged on the right.}
\end{figure}

We observed CK Vul with the VLA on 4 April 2005. The flux was 1.46 $\pm$
0.05~mJy at 5~GHz in the B configuration, the diameter is less than
0.5~arcsec. The emission is placed at the center of the observed nebulosities
(see Fig.~1). No H$\alpha$+[N\,{\sc ii}] counterpart of the radio emission is
known. This could imply that there is significant circumstellar extinction or
that the radio emission is non-thermal. Assuming the radio emission were
optically thick and the filling factor unity would yield $\Theta$ =
0.1~arcsec. Hence the upper limit for the angular extent implies that the
optical depth at 5~GHz must be at least 0.04. Evans et al. (2002) reported on
observations of the infrared spectral energy distribution of CK Vul. The
observations are consistent with a two-component model: a bigger and cooler
shell (25~K) with silicate grains, and a carbon-rich shell with a temperature
of 550~K and very small diameter. Based on angular extent, the latter shell
may be associated with the radio emission, but the evidence for this is not
conclusive. The very compact nature suggests that this is a circumstellar
disk. The slope of the SED between 450~$\mu$m and 850~$\mu$m is inconsistent
with the dust emission model, suggesting that non-thermal radio emission is
present at 850~$\mu$m.

Comparing the 2004 IPHAS H$\alpha$ image with the 1991 observations of Naylor
et al. (1992) reveals that knots 4 \& 5 identified by Shara et al. (1985) have
moved away from the central radio position at a projected velocity of roughly
100 and 200 km\,s$^{-1}$ respectively, assuming a distance of 550~pc (Fig.~1).
The arrows in Fig.~1 originate from the radio position and indicate that the
knots are indeed moving away radially from this point. The rate of
displacement of the knots is consistent with an origin during the 1670 event,
as proposed by Shara \etal\ (1985). On the IPHAS H$\alpha$ image we also
identified a larger bipolar structure centered on the radio position. Future
observations should clarify whether this structure is also connected to the
1670 event or if this could be the old PN.

\section{Conclusions}

\begin{itemize}
\item
The radio detection of the new ejecta of V4334 Sgr has been confirmed. The
radio source is more compact than previously thought and the suspected bipolar
morphology could not be confirmed. There is no apparent evidence for an
increase in ionization between 2004 and 2005.
\item
We obtained the first radio detection of V605 Aql at 5~GHz. The radio flux has
increased considerably since 1987. The current rate of evolution is
surprisingly steep considering the date of the VLTP event.
\item
We obtained the first radio detection of CK Vul. The radio source has no
counterpart in our H$\alpha$+[N\,{\sc ii}] image and is unresolved. The radio
emission may be non-thermal and associated with a circumstellar disk. The
H$\alpha$+[N\,{\sc ii}] knots identified by Shara \etal\ (1985) are confirmed
due to their proper motion as the ejecta from the 1670 event. We also detected
a large bipolar structure around CK Vul that is centered on the radio
position.
\end{itemize}

\acknowledgements PvH acknowledges support from the Belgian Science Policy
Office IAP grant P5/36.

\end{document}